\documentclass{article}
\usepackage{bezier,emlines2,gdgspace}
\title{Closing a Loophole in the
Case Against the Counterfactual Usage of the ABL Rule}
\author{ R.E. Kastner\thanks{rkastner@wam.umd.edu}}
\date{July 1, 1998}
\begin{document}
\maketitle
\begin{center}
{\small \em Department of Philosophy \\
University of Maryland \\
College Park, MD 20742 USA. \\}
\end{center}
\vspace{.1cm}
\begin{abstract}

A currently discussed interpretation of quantum 
theory, Time-Symmetrized Quantum Theory, makes certain claims
about the properties of systems between pre- and post-
selection measurements. These claims are based on a counterfactual
usage of the Aharonov-Bergmann-Lebowitz (ABL) rule
for calculating the probabilities of measurement outcomes
between such measurements. It has been argued by several 
authors that the counterfactual usage of the ABL rule is,
in general, incorrect. This paper examines what might
appear to be a loophole in those arguments and shows
that this apparent loophole cannot be used to support
a counterfactual interpretation of the ABL rule. It is
noted that the 
invalidity of the counterfactual usage of the ABL rule 
implies that the characterization
of those outcomes receiving probability 1 in a counterfactual
application of the rule as `elements of
reality' is, in general, unfounded.	
\end{abstract}
\large\vskip .2cm
{\bf 1. Introduction.~~ }

	Time-Symmetrized Quantum Theory (TSQT) is an
interpretation of quantum theory in which it is argued
that information propagating in a time-reversed direction
from future measurements can provide ontological information
about appropriately selected systems--information which is not 
taken into account in the standard time-asymmetric interpretations. 
Ensembles of such sustems are referred to as `pre- and post-selected
ensembles'. These are defined not only by the usual pre-selection
measurement at some time $t_1$, yielding the state
$\vert \psi_1(t_1)\rangle$, but also by a post-selection
measurement at time \hbox{$t_2 > t_1$,} yielding the state
$\vert \psi_2(t_2)\rangle$.  These two states
then comprise a two-state or `generalized state',
$\Psi = \langle \psi_2(t_2)\Vert \psi_1(t_1)\rangle$ 
(Aharonov and Vaidman 1991).

	TSQT advocates, most recently L. Vaidman,\footnote{\normalsize
cf. Vaidman (1996b) and (1998)} have made
certain claims about the properties of pre- and post-selected systems
based on a rule derived by Aharonov, Bergmann, and 
Lebowitz (1964), commonly known as `the ABL rule'. The ABL
rule gives
the probability of outcome $c_j$ out of the set of
possible outcomes $\{c_k\}$\footnote{\normalsize
These $\{c_k\}$ are, of course, the eigenvalues of $C$.}
 for a measurement of
nondegenerate observable $C$
occurring between pre- and post-selection measurements:

$$P_{ABL}(c_j) = {{|\langle\psi_2(t_2)|c_j\rangle|^2
|\langle c_j|\psi_1(t_1)\rangle|^2}\over{\sum_k
|\langle\psi_2(t_2)|c_k\rangle|^2
|\langle c_k|\psi_1(t_1)\rangle|^2}}\eqno(1)$$

 However, it has recently been argued by the author
 and others\footnote{\normalsize cf. 
Sharp and Shanks (1993),
Cohen (1995), Miller (1996), Kastner (1998)} that
certain of these claims are
based on a {\it counterfactual} interpretation of the ABL rule, 
which is, in general, incorrect;
and that, therefore, these claims cannot be maintained.
Specifically, the counterfactual usage allows the
outcome $c_j$ to be an eigenvalue of an observable
that has {\it not actually} been measured in the pre- and post-
selection of the system in question.

Kastner (1998) gives a quantitative characterization of
the two possible readings of the ABL rule, non-counterfactual
(which is the original, correct usage) and counterfactual
(the incorrect usage). It is shown therein that the counterfactual
usage involves a fundamental cotenability problem and that, in general,
counterfactual statements based on this usage are false.

However, in terms of Lewis' theory of counterfactuals (1973),
that argument implicitly assumed a similarity relation
(or at least a class of similarity relations)
obtaining between possible worlds. Therefore, one possible
avenue for evading the conclusion of the argument is to
challenge the similarity relation (SR); i.e., to find a different 
SR for which the aforementioned cotenability problem
does not arise. The purpose of this paper is to identify
the required class of similarity relations and to show
that any such SR will be one that makes the counterfactual
statement irrelevant or inapplicable to the actual world.

{\bf 2. A (Bad) Similarity Relation That Works.~~ }
\def\cpw{{\it j\/} }
\def\aw{{\it i\/} }
\def\jprime{{\it j'} }

	The type of counterfactual claim being made 
in the context of pre- and 
post-selected systems is as follows. Imagine
a system K pre-selected in state $\vert a\rangle$ at time
$t_1$ and post-selected in state $\vert b\rangle$ at time
$t_2$. We now consider possible measurements that {\it might}
have been made at an intermediate time $t$, $t_1 < t < t_2$,
but were not, in fact, made. The counterfactual claim is:
\vskip .2cm
\quad A: ``If we had performed a measurement of 
observable C on system K at
time $t$,\newline the probability of finding outcome $c_j$ would have 
been as given by the ABL rule.'' 
\vskip .2cm
\def\cf{\raisebox{3 pt}{\fbox{}}\!\! \to}
Statement A can be represented by the proposition:
\vskip .2cm \qquad \qquad { P\ $\cf$ Q \hskip 7cm (2)},
\vskip .2cm
where P=``Observable $C$ is measured''
and Q=``The probability of outcome $c_j$ is as given
by the ABL rule'', and the symbol `$\cf$' denotes
the counterfactual connective.

	Recall that Lewis' theory of counterfactuals
involves defining a similarity relation $\$ $ over a set of
possible worlds $\$_i$, in which \aw denotes the actual world.
$\$ $ determines what kinds of worlds will be considered
as `closer' or more `distant' in terms of similarity
to the actual world \aw. These worlds fall into
concentric spheres $\zeta^{(k)}_i$ which are nested in such
a way that the smallest sphere, $\zeta^{(1)}_i$, contains 
the actual world \aw and worlds more similar to \aw
than are any worlds in progressively larger spheres but which
are not
not also in $\zeta^{(1)}_i$. The greater the value of $k$,
the less similar to \aw are those worlds in $\zeta^{(k)}_i$ 
that are not also in $\zeta^{(k-1)}_i$. (See Figure 1.)
\vskip .5cm
\special{em:linewidth 0.4pt}
\unitlength 1.00mm
\linethickness{0.4pt}
\begin{picture}(109.16,145.67)
\emline{72.33}{101.27}{1}{74.03}{101.08}{2}
\emline{74.03}{101.08}{3}{75.65}{100.51}{4}
\emline{75.65}{100.51}{5}{77.10}{99.59}{6}
\emline{77.10}{99.59}{7}{78.30}{98.37}{8}
\emline{78.30}{98.37}{9}{79.20}{96.91}{10}
\emline{79.20}{96.91}{11}{79.76}{95.29}{12}
\emline{79.76}{95.29}{13}{79.93}{93.58}{14}
\emline{79.93}{93.58}{15}{79.72}{91.88}{16}
\emline{79.72}{91.88}{17}{79.13}{90.27}{18}
\emline{79.13}{90.27}{19}{78.19}{88.84}{20}
\emline{78.19}{88.84}{21}{76.96}{87.64}{22}
\emline{76.96}{87.64}{23}{75.49}{86.76}{24}
\emline{75.49}{86.76}{25}{73.87}{86.23}{26}
\emline{73.87}{86.23}{27}{72.16}{86.07}{28}
\emline{72.16}{86.07}{29}{70.46}{86.30}{30}
\emline{70.46}{86.30}{31}{68.86}{86.91}{32}
\emline{68.86}{86.91}{33}{67.43}{87.86}{34}
\emline{67.43}{87.86}{35}{66.25}{89.11}{36}
\emline{66.25}{89.11}{37}{65.38}{90.58}{38}
\emline{65.38}{90.58}{39}{64.87}{92.22}{40}
\emline{64.87}{92.22}{41}{64.73}{93.93}{42}
\emline{64.73}{93.93}{43}{64.99}{95.62}{44}
\emline{64.99}{95.62}{45}{65.61}{97.22}{46}
\emline{65.61}{97.22}{47}{66.58}{98.64}{48}
\emline{66.58}{98.64}{49}{67.84}{99.80}{50}
\emline{67.84}{99.80}{51}{69.32}{100.65}{52}
\emline{69.32}{100.65}{53}{72.33}{101.27}{54}
\emline{72.33}{115.84}{55}{76.28}{115.48}{56}
\emline{76.28}{115.48}{57}{80.10}{114.43}{58}
\emline{80.10}{114.43}{59}{83.67}{112.71}{60}
\emline{83.67}{112.71}{61}{86.89}{110.39}{62}
\emline{86.89}{110.39}{63}{89.63}{107.53}{64}
\emline{89.63}{107.53}{65}{91.82}{104.22}{66}
\emline{91.82}{104.22}{67}{93.39}{100.58}{68}
\emline{93.39}{100.58}{69}{94.28}{96.72}{70}
\emline{94.28}{96.72}{71}{94.48}{92.76}{72}
\emline{94.48}{92.76}{73}{93.96}{88.83}{74}
\emline{93.96}{88.83}{75}{92.75}{85.05}{76}
\emline{92.75}{85.05}{77}{90.89}{81.55}{78}
\emline{90.89}{81.55}{79}{88.43}{78.44}{80}
\emline{88.43}{78.44}{81}{85.46}{75.82}{82}
\emline{85.46}{75.82}{83}{82.07}{73.76}{84}
\emline{82.07}{73.76}{85}{78.37}{72.34}{86}
\emline{78.37}{72.34}{87}{74.48}{71.61}{88}
\emline{74.48}{71.61}{89}{70.51}{71.58}{90}
\emline{70.51}{71.58}{91}{66.60}{72.26}{92}
\emline{66.60}{72.26}{93}{62.88}{73.62}{94}
\emline{62.88}{73.62}{95}{59.46}{75.62}{96}
\emline{59.46}{75.62}{97}{56.45}{78.20}{98}
\emline{56.45}{78.20}{99}{53.95}{81.28}{100}
\emline{53.95}{81.28}{101}{52.04}{84.75}{102}
\emline{52.04}{84.75}{103}{50.77}{88.51}{104}
\emline{50.77}{88.51}{105}{50.20}{92.43}{106}
\emline{50.20}{92.43}{107}{50.33}{96.39}{108}
\emline{50.33}{96.39}{109}{51.17}{100.27}{110}
\emline{51.17}{100.27}{111}{52.68}{103.93}{112}
\emline{52.68}{103.93}{113}{54.83}{107.27}{114}
\emline{54.83}{107.27}{115}{57.53}{110.17}{116}
\emline{57.53}{110.17}{117}{60.71}{112.54}{118}
\emline{60.71}{112.54}{119}{64.25}{114.31}{120}
\emline{64.25}{114.31}{121}{68.06}{115.42}{122}
\emline{68.06}{115.42}{123}{72.33}{115.84}{124}
\emline{72.33}{130.50}{125}{77.77}{130.09}{126}
\emline{77.77}{130.09}{127}{83.08}{128.89}{128}
\emline{83.08}{128.89}{129}{88.16}{126.92}{130}
\emline{88.16}{126.92}{131}{92.90}{124.22}{132}
\emline{92.90}{124.22}{133}{97.18}{120.85}{134}
\emline{97.18}{120.85}{135}{100.92}{116.88}{136}
\emline{100.92}{116.88}{137}{104.03}{112.41}{138}
\emline{104.03}{112.41}{139}{106.45}{107.52}{140}
\emline{106.45}{107.52}{141}{108.12}{102.33}{142}
\emline{108.12}{102.33}{143}{109.01}{96.96}{144}
\emline{109.01}{96.96}{145}{109.09}{91.51}{146}
\emline{109.09}{91.51}{147}{108.37}{86.10}{148}
\emline{108.37}{86.10}{149}{106.86}{80.87}{150}
\emline{106.86}{80.87}{151}{104.59}{75.91}{152}
\emline{104.59}{75.91}{153}{101.62}{71.34}{154}
\emline{101.62}{71.34}{155}{98.00}{67.27}{156}
\emline{98.00}{67.27}{157}{93.82}{63.77}{158}
\emline{93.82}{63.77}{159}{89.17}{60.92}{160}
\emline{89.17}{60.92}{161}{84.15}{58.79}{162}
\emline{84.15}{58.79}{163}{78.88}{57.43}{164}
\emline{78.88}{57.43}{165}{73.46}{56.86}{166}
\emline{73.46}{56.86}{167}{68.01}{57.10}{168}
\emline{68.01}{57.10}{169}{62.66}{58.14}{170}
\emline{62.66}{58.14}{171}{57.52}{59.95}{172}
\emline{57.52}{59.95}{173}{52.71}{62.51}{174}
\emline{52.71}{62.51}{175}{48.32}{65.75}{176}
\emline{48.32}{65.75}{177}{44.46}{69.60}{178}
\emline{44.46}{69.60}{179}{41.21}{73.97}{180}
\emline{41.21}{73.97}{181}{38.65}{78.78}{182}
\emline{38.65}{78.78}{183}{36.82}{83.92}{184}
\emline{36.82}{83.92}{185}{35.77}{89.27}{186}
\emline{35.77}{89.27}{187}{35.52}{94.71}{188}
\emline{35.52}{94.71}{189}{36.08}{100.13}{190}
\emline{36.08}{100.13}{191}{37.43}{105.41}{192}
\emline{37.43}{105.41}{193}{39.54}{110.44}{194}
\emline{39.54}{110.44}{195}{42.38}{115.09}{196}
\emline{42.38}{115.09}{197}{45.87}{119.28}{198}
\emline{45.87}{119.28}{199}{49.94}{122.90}{200}
\emline{49.94}{122.90}{201}{54.50}{125.89}{202}
\emline{54.50}{125.89}{203}{59.45}{128.17}{204}
\emline{59.45}{128.17}{205}{64.68}{129.69}{206}
\emline{64.68}{129.69}{207}{72.33}{130.50}{208}
\put(71.33,93.67){\makebox(0,0)[cc]{{\it i}}}
\put(70.33,90.00){\makebox(0,0)[cc]{$\zeta^{(1)}_i$}}
\put(66.67,77.67){\makebox(0,0)[cc]{$\zeta^{(2)}_i$}}
\put(76.33,63.00){\makebox(0,0)[cc]{$\zeta^{(3)}_i$}}
\put(72.00,145.67){\makebox(0,0)[cc]{Figure 1}}
\put(22.33,119.33){\makebox(0,0)[cc]{${\$_i}$}}
\put(73.67,94.00){\makebox(0,0)[cc]{.}}
\put(73.00,42.67){\makebox(0,0)[cc]
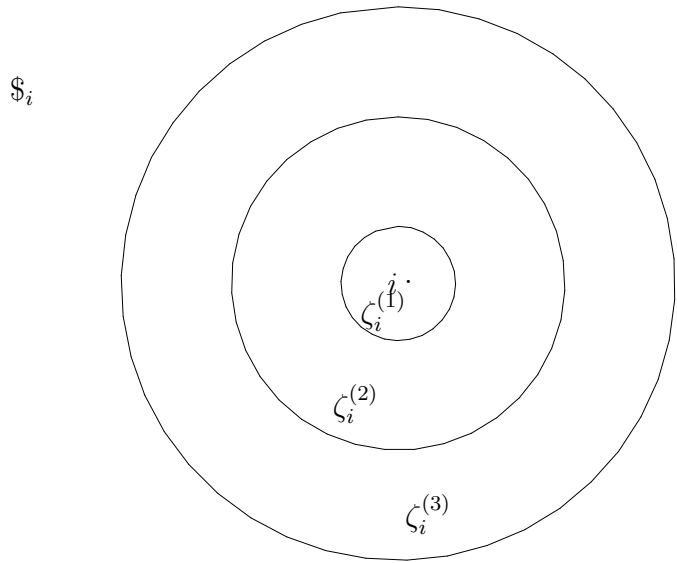
{Figure 1. The nested spheres of possible worlds in Lewis' theory of counterfactuals.}}
\end{picture}

	Further, a proposition $\psi$ is defined to be `cotenable' 
with another proposition $\phi$ at \aw iff $\psi$ holds throughout
some $\phi$-permitting sphere in $\$_i$. That is, given
a sphere $\zeta^{(k)}_i$ in which $\phi$ holds for at least
one possible world, $\psi$ holds at {\it all}
 worlds in $\zeta^{(k)}_i$. (See Figure 2.)
\vskip 1cm 
\special{em:linewidth 0.4pt}
\unitlength 1.00mm
\linethickness{0.4pt}
\begin{picture}(121.67,144.67)
\emline{73.33}{97.44}{1}{75.19}{97.23}{2}
\emline{75.19}{97.23}{3}{76.97}{96.62}{4}
\emline{76.97}{96.62}{5}{78.56}{95.63}{6}
\emline{78.56}{95.63}{7}{79.89}{94.31}{8}
\emline{79.89}{94.31}{9}{80.90}{92.73}{10}
\emline{80.90}{92.73}{11}{81.54}{90.96}{12}
\emline{81.54}{90.96}{13}{81.77}{89.10}{14}
\emline{81.77}{89.10}{15}{81.58}{87.24}{16}
\emline{81.58}{87.24}{17}{80.99}{85.46}{18}
\emline{80.99}{85.46}{19}{80.02}{83.85}{20}
\emline{80.02}{83.85}{21}{78.72}{82.50}{22}
\emline{78.72}{82.50}{23}{77.15}{81.47}{24}
\emline{77.15}{81.47}{25}{75.39}{80.82}{26}
\emline{75.39}{80.82}{27}{73.54}{80.56}{28}
\emline{73.54}{80.56}{29}{71.67}{80.73}{30}
\emline{71.67}{80.73}{31}{69.88}{81.30}{32}
\emline{69.88}{81.30}{33}{68.27}{82.25}{34}
\emline{68.27}{82.25}{35}{66.90}{83.53}{36}
\emline{66.90}{83.53}{37}{65.85}{85.09}{38}
\emline{65.85}{85.09}{39}{65.17}{86.84}{40}
\emline{65.17}{86.84}{41}{64.90}{88.69}{42}
\emline{64.90}{88.69}{43}{65.04}{90.56}{44}
\emline{65.04}{90.56}{45}{65.58}{92.35}{46}
\emline{65.58}{92.35}{47}{66.52}{93.98}{48}
\emline{66.52}{93.98}{49}{67.78}{95.36}{50}
\emline{67.78}{95.36}{51}{69.33}{96.43}{52}
\emline{69.33}{96.43}{53}{71.07}{97.13}{54}
\emline{71.07}{97.13}{55}{73.33}{97.44}{56}
\emline{73.33}{108.24}{57}{76.95}{107.90}{58}
\emline{76.95}{107.90}{59}{80.44}{106.90}{60}
\emline{80.44}{106.90}{61}{83.69}{105.27}{62}
\emline{83.69}{105.27}{63}{86.58}{103.07}{64}
\emline{86.58}{103.07}{65}{89.01}{100.37}{66}
\emline{89.01}{100.37}{67}{90.91}{97.27}{68}
\emline{90.91}{97.27}{69}{92.19}{93.87}{70}
\emline{92.19}{93.87}{71}{92.83}{90.29}{72}
\emline{92.83}{90.29}{73}{92.79}{86.66}{74}
\emline{92.79}{86.66}{75}{92.08}{83.09}{76}
\emline{92.08}{83.09}{77}{90.73}{79.72}{78}
\emline{90.73}{79.72}{79}{88.77}{76.66}{80}
\emline{88.77}{76.66}{81}{86.29}{74.01}{82}
\emline{86.29}{74.01}{83}{83.35}{71.87}{84}
\emline{83.35}{71.87}{85}{80.07}{70.30}{86}
\emline{80.07}{70.30}{87}{76.56}{69.37}{88}
\emline{76.56}{69.37}{89}{72.93}{69.11}{90}
\emline{72.93}{69.11}{91}{69.32}{69.52}{92}
\emline{69.32}{69.52}{93}{65.85}{70.59}{94}
\emline{65.85}{70.59}{95}{62.64}{72.29}{96}
\emline{62.64}{72.29}{97}{59.79}{74.55}{98}
\emline{59.79}{74.55}{99}{57.41}{77.29}{100}
\emline{57.41}{77.29}{101}{55.58}{80.43}{102}
\emline{55.58}{80.43}{103}{54.37}{83.86}{104}
\emline{54.37}{83.86}{105}{53.80}{87.44}{106}
\emline{53.80}{87.44}{107}{53.91}{91.08}{108}
\emline{53.91}{91.08}{109}{54.69}{94.63}{110}
\emline{54.69}{94.63}{111}{56.12}{97.97}{112}
\emline{56.12}{97.97}{113}{58.13}{100.99}{114}
\emline{58.13}{100.99}{115}{60.67}{103.59}{116}
\emline{60.67}{103.59}{117}{63.65}{105.67}{118}
\emline{63.65}{105.67}{119}{66.96}{107.17}{120}
\emline{66.96}{107.17}{121}{73.33}{108.24}{122}
\emline{73.33}{128.67}{123}{79.00}{128.26}{124}
\emline{79.00}{128.26}{125}{84.55}{127.05}{126}
\emline{84.55}{127.05}{127}{89.87}{125.06}{128}
\emline{89.87}{125.06}{129}{94.85}{122.33}{130}
\emline{94.85}{122.33}{131}{99.39}{118.91}{132}
\emline{99.39}{118.91}{133}{103.39}{114.88}{134}
\emline{103.39}{114.88}{135}{106.78}{110.32}{136}
\emline{106.78}{110.32}{137}{109.49}{105.32}{138}
\emline{109.49}{105.32}{139}{111.45}{99.99}{140}
\emline{111.45}{99.99}{141}{112.63}{94.43}{142}
\emline{112.63}{94.43}{143}{113.00}{88.77}{144}
\emline{113.00}{88.77}{145}{112.56}{83.10}{146}
\emline{112.56}{83.10}{147}{111.31}{77.56}{148}
\emline{111.31}{77.56}{149}{109.29}{72.25}{150}
\emline{109.29}{72.25}{151}{106.53}{67.28}{152}
\emline{106.53}{67.28}{153}{103.09}{62.76}{154}
\emline{103.09}{62.76}{155}{99.03}{58.78}{156}
\emline{99.03}{58.78}{157}{94.45}{55.42}{158}
\emline{94.45}{55.42}{159}{89.44}{52.75}{160}
\emline{89.44}{52.75}{161}{84.10}{50.82}{162}
\emline{84.10}{50.82}{163}{78.53}{49.67}{164}
\emline{78.53}{49.67}{165}{72.86}{49.33}{166}
\emline{72.86}{49.33}{167}{67.20}{49.81}{168}
\emline{67.20}{49.81}{169}{61.66}{51.08}{170}
\emline{61.66}{51.08}{171}{56.37}{53.14}{172}
\emline{56.37}{53.14}{173}{51.42}{55.93}{174}
\emline{51.42}{55.93}{175}{46.92}{59.40}{176}
\emline{46.92}{59.40}{177}{42.96}{63.48}{178}
\emline{42.96}{63.48}{179}{39.63}{68.08}{180}
\emline{39.63}{68.08}{181}{36.98}{73.10}{182}
\emline{36.98}{73.10}{183}{35.09}{78.46}{184}
\emline{35.09}{78.46}{185}{33.97}{84.03}{186}
\emline{33.97}{84.03}{187}{33.67}{89.70}{188}
\emline{33.67}{89.70}{189}{34.17}{95.36}{190}
\emline{34.17}{95.36}{191}{35.48}{100.89}{192}
\emline{35.48}{100.89}{193}{37.57}{106.17}{194}
\emline{37.57}{106.17}{195}{40.39}{111.11}{196}
\emline{40.39}{111.11}{197}{43.89}{115.59}{198}
\emline{43.89}{115.59}{199}{47.99}{119.52}{200}
\emline{47.99}{119.52}{201}{52.61}{122.83}{202}
\emline{52.61}{122.83}{203}{57.65}{125.44}{204}
\emline{57.65}{125.44}{205}{63.02}{127.31}{206}
\emline{63.02}{127.31}{207}{73.33}{128.67}{208}
\emline{87.00}{40.67}{209}{86.40}{43.65}{210}
\emline{86.40}{43.65}{211}{85.97}{46.51}{212}
\emline{85.97}{46.51}{213}{85.68}{49.25}{214}
\emline{85.68}{49.25}{215}{85.56}{51.87}{216}
\emline{85.56}{51.87}{217}{85.60}{54.37}{218}
\emline{85.60}{54.37}{219}{85.79}{56.76}{220}
\emline{85.79}{56.76}{221}{86.14}{59.03}{222}
\emline{86.14}{59.03}{223}{86.64}{61.18}{224}
\emline{86.64}{61.18}{225}{87.31}{63.21}{226}
\emline{87.31}{63.21}{227}{88.13}{65.12}{228}
\emline{88.13}{65.12}{229}{89.11}{66.92}{230}
\emline{89.11}{66.92}{231}{90.25}{68.59}{232}
\emline{90.25}{68.59}{233}{91.54}{70.15}{234}
\emline{91.54}{70.15}{235}{93.00}{71.59}{236}
\emline{93.00}{71.59}{237}{94.61}{72.92}{238}
\emline{94.61}{72.92}{239}{96.37}{74.12}{240}
\emline{96.37}{74.12}{241}{98.30}{75.21}{242}
\emline{98.30}{75.21}{243}{100.38}{76.17}{244}
\emline{100.38}{76.17}{245}{102.62}{77.02}{246}
\emline{102.62}{77.02}{247}{105.02}{77.76}{248}
\emline{105.02}{77.76}{249}{107.58}{78.37}{250}
\emline{107.58}{78.37}{251}{110.29}{78.87}{252}
\emline{110.29}{78.87}{253}{113.16}{79.24}{254}
\emline{113.16}{79.24}{255}{116.19}{79.50}{256}
\emline{116.19}{79.50}{257}{121.67}{79.67}{258}
\put(91.67,61.33){\makebox(0,0)[cc]{$\phi$}}
\put(78.00,85.67){\makebox(0,0)[cc]{$\psi$}}
\put(58.67,80.33){\makebox(0,0)[cc]{$\psi$}}
\put(54.33,61.00){\makebox(0,0)[cc]{$\psi$}}
\put(70.00,144.67){\makebox(0,0)[cc]{Figure 2}}
\put(71.33,91.00){\makebox(0,0)[cc]{{\it i}.}}
\put(71.33,25.00){\makebox(0,0)[cc]
{Figure 2. The proposition $\psi$ is cotenable with $\phi$ at {\it i} because}}
\put(71.00,20.67){\makebox(0,0)[cc]
{it holds throughout the $\phi$-permitting sphere $\zeta^{(3)}_i$.}}
\put(44.00,95.33){\makebox(0,0)[cc]{$\zeta^{(3)}_i$}}
\end{picture}

Now, according to Lewis' theory, a 
proposition of the form P\ $\cf$ Q is true at {\it i}  iff $P$ and
some auxiliary premise X, cotenable with P at {\it i},
 logically imply Q (Lewis 1973, p. 57). It was shown in
Kastner (1998) that the TSQT counterfactual claim (2) fails
to fulfill this condition because the necessary auxiliary
premise--essentially the requirement that the system
in question have the same generalized state as in world \aw
--is not cotenable with the antecedent P under a natural
similarity relation. (See Figure 3.)
\special{em:linewidth 0.4pt}
\unitlength 1.00mm
\linethickness{0.4pt}
\begin{picture}(126.67,146.67)
\emline{74.00}{100.69}{1}{76.15}{100.46}{2}
\emline{76.15}{100.46}{3}{78.21}{99.76}{4}
\emline{78.21}{99.76}{5}{80.06}{98.65}{6}
\emline{80.06}{98.65}{7}{81.63}{97.16}{8}
\emline{81.63}{97.16}{9}{82.85}{95.37}{10}
\emline{82.85}{95.37}{11}{83.65}{93.35}{12}
\emline{83.65}{93.35}{13}{84.01}{91.22}{14}
\emline{84.01}{91.22}{15}{83.89}{89.05}{16}
\emline{83.89}{89.05}{17}{83.31}{86.97}{18}
\emline{83.31}{86.97}{19}{82.30}{85.05}{20}
\emline{82.30}{85.05}{21}{80.90}{83.40}{22}
\emline{80.90}{83.40}{23}{79.17}{82.09}{24}
\emline{79.17}{82.09}{25}{77.21}{81.18}{26}
\emline{77.21}{81.18}{27}{75.09}{80.71}{28}
\emline{75.09}{80.71}{29}{72.93}{80.71}{30}
\emline{72.93}{80.71}{31}{70.81}{81.17}{32}
\emline{70.81}{81.17}{33}{68.85}{82.08}{34}
\emline{68.85}{82.08}{35}{67.12}{83.39}{36}
\emline{67.12}{83.39}{37}{65.71}{85.04}{38}
\emline{65.71}{85.04}{39}{64.70}{86.95}{40}
\emline{64.70}{86.95}{41}{64.11}{89.03}{42}
\emline{64.11}{89.03}{43}{63.99}{91.20}{44}
\emline{63.99}{91.20}{45}{64.34}{93.33}{46}
\emline{64.34}{93.33}{47}{65.14}{95.35}{48}
\emline{65.14}{95.35}{49}{66.35}{97.14}{50}
\emline{66.35}{97.14}{51}{67.92}{98.64}{52}
\emline{67.92}{98.64}{53}{69.77}{99.76}{54}
\emline{69.77}{99.76}{55}{71.83}{100.45}{56}
\emline{71.83}{100.45}{57}{74.00}{100.69}{58}
\emline{73.67}{135.04}{59}{79.69}{134.63}{60}
\emline{79.69}{134.63}{61}{85.59}{133.40}{62}
\emline{85.59}{133.40}{63}{91.28}{131.39}{64}
\emline{91.28}{131.39}{65}{96.64}{128.63}{66}
\emline{96.64}{128.63}{67}{101.58}{125.16}{68}
\emline{101.58}{125.16}{69}{106.00}{121.06}{70}
\emline{106.00}{121.06}{71}{109.82}{116.39}{72}
\emline{109.82}{116.39}{73}{112.97}{111.25}{74}
\emline{112.97}{111.25}{75}{115.40}{105.73}{76}
\emline{115.40}{105.73}{77}{117.06}{99.93}{78}
\emline{117.06}{99.93}{79}{117.91}{93.96}{80}
\emline{117.91}{93.96}{81}{117.95}{87.93}{82}
\emline{117.95}{87.93}{83}{117.17}{81.95}{84}
\emline{117.17}{81.95}{85}{115.58}{76.13}{86}
\emline{115.58}{76.13}{87}{113.22}{70.58}{88}
\emline{113.22}{70.58}{89}{110.13}{65.40}{90}
\emline{110.13}{65.40}{91}{106.37}{60.69}{92}
\emline{106.37}{60.69}{93}{102.00}{56.53}{94}
\emline{102.00}{56.53}{95}{97.11}{53.00}{96}
\emline{97.11}{53.00}{97}{91.78}{50.17}{98}
\emline{91.78}{50.17}{99}{86.12}{48.09}{100}
\emline{86.12}{48.09}{101}{80.23}{46.79}{102}
\emline{80.23}{46.79}{103}{74.22}{46.31}{104}
\emline{74.22}{46.31}{105}{68.20}{46.64}{106}
\emline{68.20}{46.64}{107}{62.28}{47.79}{108}
\emline{62.28}{47.79}{109}{56.57}{49.73}{110}
\emline{56.57}{49.73}{111}{51.17}{52.43}{112}
\emline{51.17}{52.43}{113}{46.20}{55.84}{114}
\emline{46.20}{55.84}{115}{41.72}{59.88}{116}
\emline{41.72}{59.88}{117}{37.84}{64.50}{118}
\emline{37.84}{64.50}{119}{34.63}{69.60}{120}
\emline{34.63}{69.60}{121}{32.13}{75.09}{122}
\emline{32.13}{75.09}{123}{30.40}{80.87}{124}
\emline{30.40}{80.87}{125}{29.47}{86.83}{126}
\emline{29.47}{86.83}{127}{29.36}{92.86}{128}
\emline{29.36}{92.86}{129}{30.07}{98.85}{130}
\emline{30.07}{98.85}{131}{31.58}{104.69}{132}
\emline{31.58}{104.69}{133}{33.87}{110.27}{134}
\emline{33.87}{110.27}{135}{36.89}{115.49}{136}
\emline{36.89}{115.49}{137}{40.60}{120.24}{138}
\emline{40.60}{120.24}{139}{44.92}{124.46}{140}
\emline{44.92}{124.46}{141}{49.77}{128.04}{142}
\emline{49.77}{128.04}{143}{55.06}{130.94}{144}
\emline{55.06}{130.94}{145}{60.69}{133.09}{146}
\emline{60.69}{133.09}{147}{66.56}{134.46}{148}
\emline{66.56}{134.46}{149}{73.67}{135.04}{150}
\emline{82.33}{34.67}{151}{81.20}{38.47}{152}
\emline{81.20}{38.47}{153}{80.24}{42.12}{154}
\emline{80.24}{42.12}{155}{79.44}{45.64}{156}
\emline{79.44}{45.64}{157}{78.80}{49.02}{158}
\emline{78.80}{49.02}{159}{78.32}{52.26}{160}
\emline{78.32}{52.26}{161}{78.00}{55.35}{162}
\emline{78.00}{55.35}{163}{77.85}{58.31}{164}
\emline{77.85}{58.31}{165}{77.85}{61.13}{166}
\emline{77.85}{61.13}{167}{78.02}{63.80}{168}
\emline{78.02}{63.80}{169}{78.35}{66.34}{170}
\emline{78.35}{66.34}{171}{78.84}{68.74}{172}
\emline{78.84}{68.74}{173}{79.49}{70.99}{174}
\emline{79.49}{70.99}{175}{80.30}{73.11}{176}
\emline{80.30}{73.11}{177}{81.27}{75.09}{178}
\emline{81.27}{75.09}{179}{82.41}{76.92}{180}
\emline{82.41}{76.92}{181}{83.71}{78.62}{182}
\emline{83.71}{78.62}{183}{85.17}{80.18}{184}
\emline{85.17}{80.18}{185}{86.79}{81.59}{186}
\emline{86.79}{81.59}{187}{88.57}{82.87}{188}
\emline{88.57}{82.87}{189}{90.51}{84.00}{190}
\emline{90.51}{84.00}{191}{92.61}{85.00}{192}
\emline{92.61}{85.00}{193}{94.88}{85.86}{194}
\emline{94.88}{85.86}{195}{97.31}{86.57}{196}
\emline{97.31}{86.57}{197}{99.90}{87.15}{198}
\emline{99.90}{87.15}{199}{102.65}{87.58}{200}
\emline{102.65}{87.58}{201}{105.56}{87.88}{202}
\emline{105.56}{87.88}{203}{108.63}{88.03}{204}
\emline{108.63}{88.03}{205}{111.87}{88.05}{206}
\emline{111.87}{88.05}{207}{115.26}{87.92}{208}
\emline{115.26}{87.92}{209}{118.82}{87.66}{210}
\emline{118.82}{87.66}{211}{122.54}{87.25}{212}
\emline{122.54}{87.25}{213}{126.67}{86.67}{214}
\put(84.33,40.33){\makebox(0,0)[cc]{P}}
\put(73.33,91.67){\makebox(0,0)[cc]{{\it i}.}}
\put(73.67,86.67){\makebox(0,0)[cc]{$\neg$P\&T}}
\put(60.33,77.00){\makebox(0,0)[cc]{T$\vee\neg$T}}
\put(96.66,69.66){\makebox(0,0)[cc]{P\&($T\vee\neg$T)}}
\put(71.67,146.67){\makebox(0,0)[cc]{Figure 3}}
\put(73.33,97.00){\makebox(0,0)[cc]{$\zeta^{(1)}_i$}}
\put(64.67,112.33){\makebox(0,0)[cc]{$\zeta^{(2)}_i$}}
\put(74.33,21.00){\makebox(0,0)[cc]
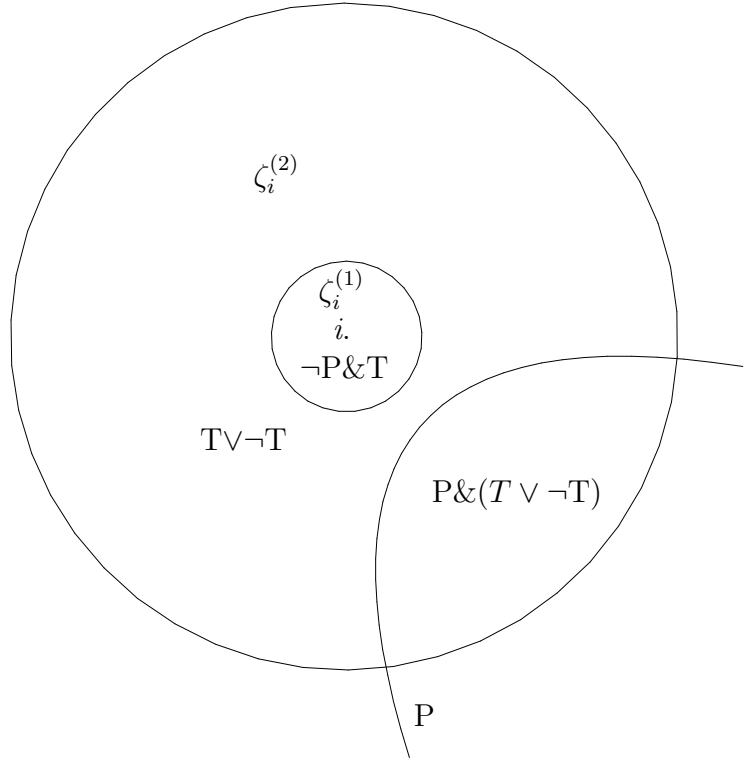
{Figure 3. A natural similarity relation equivalent to that presented in Kastner (1998).}}
\end{picture}

	The SR shown in Figure 3 is one in which it is assumed
that the occurrence of a measurement at time $t$, denoted
by the proposition P, affects the possible outcomes
of the post-selection measurement at time $t_2$. Therefore,
a system K that has generalized state $\Psi$ in \aw, where
$\neg$P holds, {\it might not} necessarily have that same state
in another world {\it m} in which P holds. In Figure 3, the proposition
T states that system K has the same
generalized state as in \aw. In the second sphere $\zeta^{(2)}_i$,
the occurrence of possible measurements at time $t$ (P stating
that one particular kind of measurement has been performed)
results in an ambiguous result for the post-selection state
for K and therefore there may be possible worlds 
in this sphere in which T holds and other worlds in which 
it does not. Since, in order for T to be
cotenable with P, it must hold throughout a P-permitting
sphere--i.e. at {\it all} worlds in that sphere--
proposition T is not cotenable
with P under this similarity relation.

	It was also shown in Kastner (1998) that Vaidman's definition
of a `closest possible world' \cpw (equivalently a choice
of ${\$}$ for which the counterfactual usage of the ABL rule
was correct) was untenable because that closest possible
world did not, in general, exist. That definition, presented
in Vaidman (1996a),
stipulated that the closest possible world was one in which
all measurements, excluding the 
intermediate measurement asserted
by P, have the same outcomes as in \aw.
\vskip .2cm

	However, there {\it is} a way to define a `closest possible world' 
\cpw, equivalently a similarity relation, in such a way that 
\cpw in general does exist and so 
that there is, at least formally, no problem with cotenability 
as described in Kastner (1998).  Call this similarity relation 
Z, and define it as follows:

Given a system K in the actual world \aw 
(where no intervening measurement has been performed, i.e.
$\neg P$ holds) with two-state vector 
$\Psi = \langle b||a\rangle$, \cpw is the world in which the 
following conjunction holds:
\vskip .2cm
\qquad\qquad	P \& T \hskip 7cm (3)
\vskip .2cm
where T, as in the preceding discussion of Figure 3,
 is the proposition: ``K has two-state vector $\Psi$.''

	In terms of Lewis' system of concentric spheres, 
Z is the configuration in which all worlds in which 
P \& T holds are assigned to the smallest sphere 
 in which P holds.    
In other words, T is the auxiliary proposition that, 
together with P, implies the consequent Q.
 Since T holds throughout a P-permitting sphere, it is 
cotenable with P under Z (see figure 4).
\special{em:linewidth 0.4pt}
\unitlength 1.00mm
\linethickness{0.4pt}
\begin{picture}(136.00,147.00)
\emline{74.00}{124.10}{1}{79.40}{123.70}{2}
\emline{79.40}{123.70}{3}{84.68}{122.50}{4}
\emline{84.68}{122.50}{5}{89.73}{120.53}{6}
\emline{89.73}{120.53}{7}{94.43}{117.83}{8}
\emline{94.43}{117.83}{9}{98.68}{114.47}{10}
\emline{98.68}{114.47}{11}{102.38}{110.51}{12}
\emline{102.38}{110.51}{13}{105.45}{106.05}{14}
\emline{105.45}{106.05}{15}{107.83}{101.19}{16}
\emline{107.83}{101.19}{17}{109.46}{96.02}{18}
\emline{109.46}{96.02}{19}{110.31}{90.67}{20}
\emline{110.31}{90.67}{21}{110.35}{85.25}{22}
\emline{110.35}{85.25}{23}{109.59}{79.89}{24}
\emline{109.59}{79.89}{25}{108.04}{74.70}{26}
\emline{108.04}{74.70}{27}{105.74}{69.79}{28}
\emline{105.74}{69.79}{29}{102.74}{65.28}{30}
\emline{102.74}{65.28}{31}{99.10}{61.27}{32}
\emline{99.10}{61.27}{33}{94.91}{57.84}{34}
\emline{94.91}{57.84}{35}{90.26}{55.07}{36}
\emline{90.26}{55.07}{37}{85.24}{53.02}{38}
\emline{85.24}{53.02}{39}{79.98}{51.73}{40}
\emline{79.98}{51.73}{41}{74.58}{51.24}{42}
\emline{74.58}{51.24}{43}{69.17}{51.56}{44}
\emline{69.17}{51.56}{45}{63.87}{52.68}{46}
\emline{63.87}{52.68}{47}{58.80}{54.56}{48}
\emline{58.80}{54.56}{49}{54.06}{57.18}{50}
\emline{54.06}{57.18}{51}{49.76}{60.48}{52}
\emline{49.76}{60.48}{53}{45.99}{64.38}{54}
\emline{45.99}{64.38}{55}{42.85}{68.79}{56}
\emline{42.85}{68.79}{57}{40.39}{73.61}{58}
\emline{40.39}{73.61}{59}{38.68}{78.75}{60}
\emline{38.68}{78.75}{61}{37.75}{84.09}{62}
\emline{37.75}{84.09}{63}{37.62}{89.51}{64}
\emline{37.62}{89.51}{65}{38.29}{94.88}{66}
\emline{38.29}{94.88}{67}{39.75}{100.10}{68}
\emline{39.75}{100.10}{69}{41.98}{105.04}{70}
\emline{41.98}{105.04}{71}{44.91}{109.59}{72}
\emline{44.91}{109.59}{73}{48.48}{113.67}{74}
\emline{48.48}{113.67}{75}{52.62}{117.16}{76}
\emline{52.62}{117.16}{77}{57.23}{120.01}{78}
\emline{57.23}{120.01}{79}{62.21}{122.14}{80}
\emline{62.21}{122.14}{81}{67.45}{123.51}{82}
\emline{67.45}{123.51}{83}{74.00}{124.10}{84}
\put(85.33,31.33){\makebox(0,0)[cc]{P}}
\put(73.67,88.00){\makebox(0,0)[cc]{{\it i}.}}
\put(73.33,83.67){\makebox(0,0)[cc]{$\neg$P\&T}}
\put(54.67,72.33){\makebox(0,0)[cc]{T}}
\put(93.66,68.00){\makebox(0,0)[cc]{P\&T}}
\put(50.66,49.67){\makebox(0,0)[cc]{$\neg$T}}
\put(103.00,52.00){\makebox(0,0)[cc]{P\&$\neg$T}}
\put(72.00,147.00){\makebox(0,0)[cc]{Figure 4}}
\put(21.00,115.33){\makebox(0,0)[cc]{\Large Z}}
\emline{73.67}{99.90}{85}{76.15}{99.64}{86}
\emline{76.15}{99.64}{87}{78.52}{98.87}{88}
\emline{78.52}{98.87}{89}{80.67}{97.62}{90}
\emline{80.67}{97.62}{91}{82.52}{95.95}{92}
\emline{82.52}{95.95}{93}{83.99}{93.93}{94}
\emline{83.99}{93.93}{95}{85.00}{91.65}{96}
\emline{85.00}{91.65}{97}{85.51}{89.21}{98}
\emline{85.51}{89.21}{99}{85.50}{86.72}{100}
\emline{85.50}{86.72}{101}{84.98}{84.29}{102}
\emline{84.98}{84.29}{103}{83.95}{82.01}{104}
\emline{83.95}{82.01}{105}{82.48}{80.00}{106}
\emline{82.48}{80.00}{107}{80.62}{78.34}{108}
\emline{80.62}{78.34}{109}{78.46}{77.11}{110}
\emline{78.46}{77.11}{111}{76.09}{76.35}{112}
\emline{76.09}{76.35}{113}{73.61}{76.10}{114}
\emline{73.61}{76.10}{115}{71.13}{76.37}{116}
\emline{71.13}{76.37}{117}{68.77}{77.16}{118}
\emline{68.77}{77.16}{119}{66.61}{78.42}{120}
\emline{66.61}{78.42}{121}{64.77}{80.10}{122}
\emline{64.77}{80.10}{123}{63.32}{82.12}{124}
\emline{63.32}{82.12}{125}{62.33}{84.41}{126}
\emline{62.33}{84.41}{127}{61.83}{86.85}{128}
\emline{61.83}{86.85}{129}{61.85}{89.34}{130}
\emline{61.85}{89.34}{131}{62.38}{91.77}{132}
\emline{62.38}{91.77}{133}{63.42}{94.04}{134}
\emline{63.42}{94.04}{135}{64.90}{96.04}{136}
\emline{64.90}{96.04}{137}{66.77}{97.69}{138}
\emline{66.77}{97.69}{139}{68.94}{98.92}{140}
\emline{68.94}{98.92}{141}{73.67}{99.90}{142}
\emline{73.67}{139.57}{143}{80.20}{139.16}{144}
\emline{80.20}{139.16}{145}{86.63}{137.93}{146}
\emline{86.63}{137.93}{147}{92.86}{135.91}{148}
\emline{92.86}{135.91}{149}{98.78}{133.13}{150}
\emline{98.78}{133.13}{151}{104.31}{129.63}{152}
\emline{104.31}{129.63}{153}{109.36}{125.47}{154}
\emline{109.36}{125.47}{155}{113.85}{120.71}{156}
\emline{113.85}{120.71}{157}{117.71}{115.42}{158}
\emline{117.71}{115.42}{159}{120.87}{109.70}{160}
\emline{120.87}{109.70}{161}{123.30}{103.62}{162}
\emline{123.30}{103.62}{163}{124.95}{97.28}{164}
\emline{124.95}{97.28}{165}{125.79}{90.79}{166}
\emline{125.79}{90.79}{167}{125.81}{84.25}{168}
\emline{125.81}{84.25}{169}{125.02}{77.75}{170}
\emline{125.02}{77.75}{171}{123.42}{71.41}{172}
\emline{123.42}{71.41}{173}{121.04}{65.31}{174}
\emline{121.04}{65.31}{175}{117.91}{59.56}{176}
\emline{117.91}{59.56}{177}{114.09}{54.25}{178}
\emline{114.09}{54.25}{179}{109.64}{49.45}{180}
\emline{109.64}{49.45}{181}{104.62}{45.25}{182}
\emline{104.62}{45.25}{183}{99.12}{41.71}{184}
\emline{99.12}{41.71}{185}{93.21}{38.89}{186}
\emline{93.21}{38.89}{187}{87.00}{36.83}{188}
\emline{87.00}{36.83}{189}{80.58}{35.55}{190}
\emline{80.58}{35.55}{191}{74.05}{35.10}{192}
\emline{74.05}{35.10}{193}{67.52}{35.46}{194}
\emline{67.52}{35.46}{195}{61.08}{36.63}{196}
\emline{61.08}{36.63}{197}{54.84}{38.61}{198}
\emline{54.84}{38.61}{199}{48.90}{41.34}{200}
\emline{48.90}{41.34}{201}{43.34}{44.80}{202}
\emline{43.34}{44.80}{203}{38.26}{48.93}{204}
\emline{38.26}{48.93}{205}{33.74}{53.66}{206}
\emline{33.74}{53.66}{207}{29.84}{58.91}{208}
\emline{29.84}{58.91}{209}{26.63}{64.62}{210}
\emline{26.63}{64.62}{211}{24.16}{70.68}{212}
\emline{24.16}{70.68}{213}{22.47}{77.00}{214}
\emline{22.47}{77.00}{215}{21.58}{83.48}{216}
\emline{21.58}{83.48}{217}{21.50}{90.03}{218}
\emline{21.50}{90.03}{219}{22.25}{96.53}{220}
\emline{22.25}{96.53}{221}{23.80}{102.88}{222}
\emline{23.80}{102.88}{223}{26.14}{109.00}{224}
\emline{26.14}{109.00}{225}{29.22}{114.77}{226}
\emline{29.22}{114.77}{227}{33.00}{120.11}{228}
\emline{33.00}{120.11}{229}{37.42}{124.94}{230}
\emline{37.42}{124.94}{231}{42.41}{129.18}{232}
\emline{42.41}{129.18}{233}{47.89}{132.76}{234}
\emline{47.89}{132.76}{235}{53.77}{135.63}{236}
\emline{53.77}{135.63}{237}{59.97}{137.74}{238}
\emline{59.97}{137.74}{239}{66.38}{139.05}{240}
\emline{66.38}{139.05}{241}{73.67}{139.57}{242}
\emline{81.67}{28.00}{243}{80.91}{31.89}{244}
\emline{80.91}{31.89}{245}{80.30}{35.66}{246}
\emline{80.30}{35.66}{247}{79.85}{39.29}{248}
\emline{79.85}{39.29}{249}{79.55}{42.79}{250}
\emline{79.55}{42.79}{251}{79.40}{46.17}{252}
\emline{79.40}{46.17}{253}{79.40}{49.41}{254}
\emline{79.40}{49.41}{255}{79.56}{52.52}{256}
\emline{79.56}{52.52}{257}{79.87}{55.51}{258}
\emline{79.87}{55.51}{259}{80.33}{58.36}{260}
\emline{80.33}{58.36}{261}{80.95}{61.08}{262}
\emline{80.95}{61.08}{263}{81.72}{63.68}{264}
\emline{81.72}{63.68}{265}{82.64}{66.14}{266}
\emline{82.64}{66.14}{267}{83.72}{68.48}{268}
\emline{83.72}{68.48}{269}{84.95}{70.68}{270}
\emline{84.95}{70.68}{271}{86.33}{72.76}{272}
\emline{86.33}{72.76}{273}{87.86}{74.70}{274}
\emline{87.86}{74.70}{275}{89.55}{76.52}{276}
\emline{89.55}{76.52}{277}{91.39}{78.20}{278}
\emline{91.39}{78.20}{279}{93.38}{79.76}{280}
\emline{93.38}{79.76}{281}{95.53}{81.18}{282}
\emline{95.53}{81.18}{283}{97.82}{82.48}{284}
\emline{97.82}{82.48}{285}{100.28}{83.65}{286}
\emline{100.28}{83.65}{287}{102.88}{84.68}{288}
\emline{102.88}{84.68}{289}{105.64}{85.59}{290}
\emline{105.64}{85.59}{291}{108.55}{86.37}{292}
\emline{108.55}{86.37}{293}{111.61}{87.01}{294}
\emline{111.61}{87.01}{295}{114.83}{87.53}{296}
\emline{114.83}{87.53}{297}{118.20}{87.92}{298}
\emline{118.20}{87.92}{299}{121.72}{88.17}{300}
\emline{121.72}{88.17}{301}{125.39}{88.30}{302}
\emline{125.39}{88.30}{303}{129.22}{88.30}{304}
\emline{129.22}{88.30}{305}{136.00}{88.00}{306}
\put(103.00,80.00){\makebox(0,0)[cc]{{\it j}.}}
\put(116.67,69.00){\makebox(0,0)[cc]{{\it j'}.}}
\put(72.00,96.00){\makebox(0,0)[cc]{$\zeta^{(1)}_i$}}
\put(64.67,117.00){\makebox(0,0)[cc]{$\zeta^{(2)}_i$}}
\put(66.33,133.67){\makebox(0,0)[cc]{$\zeta^{(3)}_i$}}
\put(75.00,16.67){\makebox(0,0)[cc]
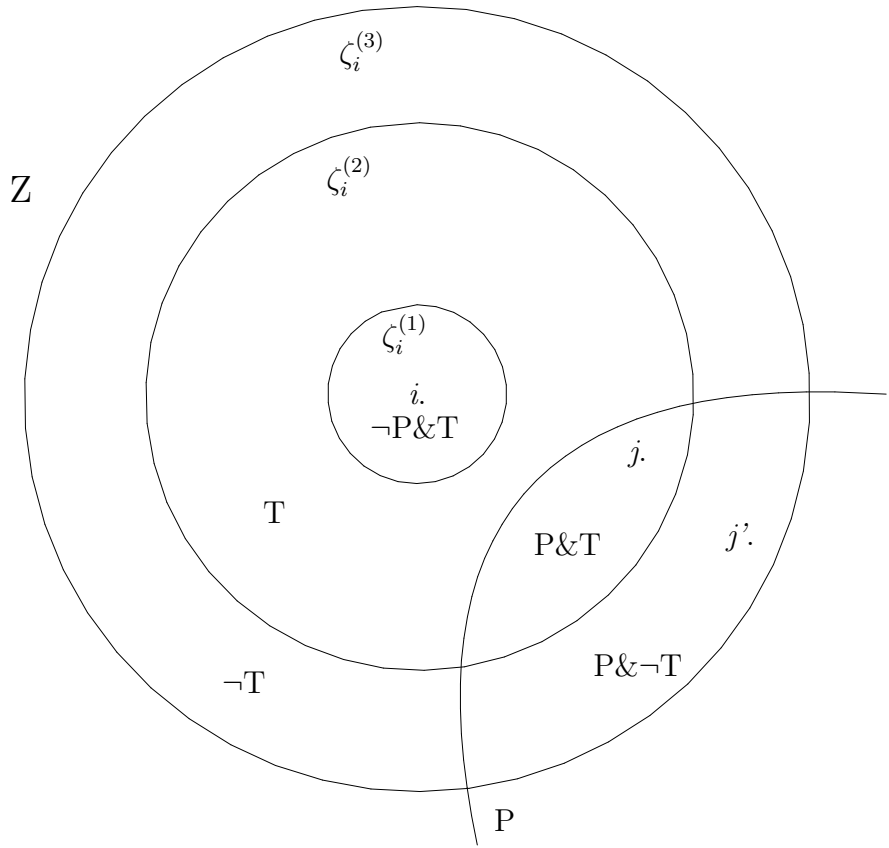
{Figure 4. The similarity relation Z that attempts to exploit the loophole.}}
\end{picture}

	This is almost the same as Vaidman's definition  
for a closest possible world, except that here the only 
measurement outcome (after time $t$) that is required to be 
the same as in \aw is the post-selection measurement outcome 
for system K (rather than requiring that {\it all} measurement 
outcomes be the same). In general, we should always be 
able to find such a world (barring cases in which the 
post-selection state  $|b\rangle$ is orthogonal to the 
state resulting from the measurement at time $t$).
 
	Similarity relation Z sidesteps the cotenability problem because,
 at least for world \cpw,  the intervening measurement is simply
{\it stipulated} not to change 
the background conditions (i.e., the pre- and post-selected states), 
asserted to hold at \cpw  by the proposition T. 

	This may be what Vaidman has in mind in his alternative definition
  for a time-symmetrized counterfactual, which 
involves `fixing' the pre- and post-selected states (Vaidman 1998).
 It was argued in Kastner (1998) that this definition is untenable
 because there is no way to accomplish this fixing requirement.
This difficulty is related to what is wrong with the  choice of Z as 
similarity relation. 

{\bf 3. Similarity defined in terms of likelihood.}
	
	In choosing a similarity relation relative to the actual 
world \aw, we have to be loyal to the physical and probabilistic 
structure of the actual world \aw. What has to be argued 
in order to justify similarity relation Z 
is that worlds assigned to the sphere $\zeta^{(2)}_i$ in which T holds 
are really more similar to \aw 
than worlds assigned to $\zeta^{(3)}_i$ in which $\neg$T holds. 
We should therefore be able to give arguments justifying the 
characterization of the `closest possible world' \cpw as more similar 
to \aw than some other P-world {\it j'} in which system K has a 
different post-selection outcome than in \aw.

The tempting argument to make here is that \cpw is objectively more similar 
to \aw than is  \jprime in virtue of K's having the 
same post-selection outcome in \cpw as in \aw; i.e.,
the fact that T holds at \cpw but not at \jprime. However, as Lewis
has pointed out (1973, pp. 52-53), what is relevant in deciding whether
the holding of T at any particular world makes that world
closer to \aw than some other world in which T does not hold,
is the {\it comparative 
possibility} of T with respect to \aw, rather than the superficial similarity 
of world \cpw to world \aw owing to the truth of T in both worlds.

The term `comparative possibility' is the notion that some propositions
are more possible than others given the relevant conditions 
obtaining in the actual world \aw. This can be seen as an intuitively
sensible measure of similarity of worlds, as Lewis shows via this 
 illustration:
``It is more possible for a dog to talk than for a stone to talk,
since some worlds with talking dogs are more like our world than is any world 
with talking stones.''

	One way to quantify this notion of comparative possibility
of propositions is through the {\it likelihood} 
of a given apparent point of similarity, say X,
  conditional on some proposition of interest P
not holding in \aw, and relevant
background conditions in \aw. In order for some world \cpw to qualify
as more similar to \aw than some other world \jprime based
on the coincidence of X in worlds \aw and \cpw but
not in world \jprime, the likelihood of X given
P (and background conditions B) must be greater than the likelihood
of $\neg$X given P\&B. Thus, we require:
\vskip .2cm
\quad Prob(X$\vert$ P\&B) $>$ Prob($\neg$X$\vert$ P\&B).\hskip 7cm (4)

	An example may serve to clarify this point. Suppose that in 
the actual world \aw, I (along with 10 million others) enter the lottery 
at time $t$ and do not win at time $t'$. Now consider a world {\it m}
 in which 
I am the only entrant at time $t$, but I {\it still}
 do not win the lottery at 
time $t'$ in {\it m}. Superficially, world {\it m}
 is much like \aw in that the outcomes 
at $t'$ are the same (I lose in both worlds). But we would not
 consider world {\it m} to be close 
to \aw under any natural similarity relation, since according to the 
laws of world \aw, the outcome of my losing (call it L) at time $t'$
 given that I am 
the only entrant is {\it impossible} (assuming the lottery managers aren't 
cheating). In term of the likelihood criterion, 
the likelihood of L conditional
on the proposition of interest in {\it m} which causes it to differ from
 \aw (i.e., the fact
that I am the only entrant) and the relevant laws in \aw
(i.e., the laws of probability) is zero. 
Thus the fact that the outcomes at time $t'$
 are the same in 
both worlds is a freak occurrence rather than a legitimate indicator 
of closeness of those worlds.

	The above example is an extreme case in which outcome L at $t'$
 in world {\it m} would be impossible according to the laws in 
world \aw. However, 
it may just be that the likelihood of the outcome or event 
under consideration as a criterion of similarity between worlds
{\it m} and \aw is less than 1 
according to the laws in \aw; rather than being impossible, the event is not 
guaranteed or possibly even farfetched to some degree. The more
the likelihood deviates from 1, the further is the
associated world from \aw in terms of similarity.

	Of particular importance is the case in which we are 
considering the counterfactual \hbox{P\ $\cf$ Q}, and {\it m}
 is a P-world in which 
some background condition or auxiliary premise T holding in \aw
 also holds in {\it m}, despite the fact that according to the laws in 
\aw (i.e., the background conditions B), the holding of P would 
make T less likely.  
Here again we have a superficial resemblance of worlds \aw
and {\it m}. 
However, the likelihood of the event T given P is less than 1. 
 Therefore the occurrence of T in both worlds is not a valid 
indicator of closeness.

	Now suppose that the likelihood of T given P and the laws 
holding in \aw is 1/2. Consider another P-world {\it m$'$}
 identical to {\it m} except for the fact that $\neg$T holds in 
{\it m$'$}.  There is no basis for saying that {\it m}
 is closer to \aw than is {\it m$'$} in virtue of the holding of T, 
because it was no more likely that T than that $\neg$T. 
By the comparative possibility  criterion, the worlds 
{\it m} and {\it m$'$} are equally close to \aw, and so belong in 
the same sphere.

	This is exactly the kind of problem we have under Z for the 
counterfactual  P\ $\cf$ Q. There is no basis for saying that 
world \cpw is more similar to \aw than world \jprime,
 i.e., that \cpw belongs in a smaller sphere than \jprime,
  if the likelihoods of T and $\neg$T  given P are the same. 
Indeed, if $\neg$T were {\it more} likely than T given P and the 
laws holding in \aw, then  \jprime would belong in a smaller 
sphere than \cpw.

	Quite simply, choosing Z as the similarity relation 
amounts to `stacking the deck.' To change the earlier
example a bit: it is like saying, ``If I were
 to enter the lottery, I would win,'' and defining the closest 
possible world as the one in which I am the only entrant.
 Under this similarity relation, the counterfactual is 
true; but it doesn't
tell me very much about the actual world.

{\bf 4. Conclusion.}

	It has been argued that an apparent loophole
in arguments in Kastner (1998) establishing that the counterfactual
usage of the ABL rule is, in general, invalid, is
not a genuine loophole. This is because the only way
to exploit the loophole is to
use a similarity relation over the set of possible worlds 
that renders the
counterfactual statement irrelevant to the actual world. 
Thus the conclusion of that paper stands: the ABL rule cannot,
in general, be used to calculate the probability of various
outcomes of an intervening measurement on a pre- and post-selected system 
if that measurement was not actually performed
on the given system. Therefore, it is erroneous to refer
to outcomes having probability 1 in a counterfactual application
of the ABL rule as being `elements of reality' of a system,
as is done in Vaidman (1996b) and (1998).
	
In his most recent paper, Vaidman (1998) suggests that
criticisms of the counterfactual usage of the ABL rule arise
from a misunderstanding of the role of time symmetry in counterfactual
statements. However, it should be noted that 
the conclusions of this paper and of Kastner (1998)
in no way depend on any assumption of time asymmetry.
\newpage
{\bf References}

\noindent Aharonov, Y, Bergmann, P.G. , and Lebowitz, J.L. (1964), 
`Time Symmetry in the Quantum Process of Measurement,' 
{\it Physical Review B 134}, 1410-16.\newline
Aharonov, Y. and Vaidman, L. (1991), `Complete Description
of a Quantum System at a Given Time,'
{\it Journal of Physics A 24}, 2315-28.\newline
Cohen, O. (1995), `Pre- and postselected quantum systems, 
counterfactual 
measurements, and consistent histories, '
{\it Physical Review A 51}, 4373-4380.\newline
Kastner, R. E. (1998), `Time-Symmetrized Quantum Theory,
Counterfactuals, and ``Advanced Action'',' quant-ph/9806002,
forthcoming in {\it Studies in History and Philosophy 
of Modern Physics}.\newline
Lewis, D. (1973), {\it Counterfactuals}, Cambridge: 
Harvard University Press. \newline 
Miller, D.J. (1996), `Realism and Time Symmetry in 
Quantum Mechanics,' {\it Physics Letters A 222}, 31.\newline
Sharp, W. and Shanks, N. (1993), `The Rise and Fall of 
Time-Symmetrized 
Quantum Mechanics,' {\it Philosophy of Science 60}, 
488-499.\newline
Vaidman, L. (1996a), `Defending Time-Symmetrized Quantum 
Theory,' preprint, quant-phys/9609007. \newline
Vaidman, L. (1996b), `Weak-Measurement Elements of Reality,' 
{\it Foundations of Physics 26}, 895-906.\newline
Vaidman, L.(1998), `Time-Symmetrized Counterfactuals
in Quantum Theory,' preprint, quant-ph/9802042.
\end{document}